\documentclass[aps,onecolumn,12pt]{revtex4-2}
\usepackage[T1]{fontenc}
\usepackage[english]{babel}
\usepackage{microtype}
\usepackage{amsmath,amssymb,amsfonts}
\usepackage{bm}
\usepackage{graphicx}
\usepackage{xcolor}

\begin{document}

\title{Wave Phenomena and Wave Equations\footnote{$1^{st}$ revised version of a lecture presented by Gerd Leuchs at the Enrico Fermi Summer School at Varenna, Course 197
'Foundations of Quantum Theory', 2016, Organizers: Ernst M Rasel,
Wolfgang P Schleich and Sabine Wölk (published in the proceedings 2019)}}

\author{Gerd Leuchs}
\affiliation{Department Physik, Friedrich-Alexander-Universität Erlangen--Nürnberg, Germany}
\affiliation{Department of Physics, University of Ottawa, Canada}
\affiliation{Max Planck Institute for the Science of Light, Germany}

\author{Mojdeh S. Najafabadi}
\affiliation{Max Planck Institute for the Science of Light, Germany}

\date{\today}
\maketitle{}

\section*{Summary}
For	any	kind of	wave phenomenon	one	can	find	ways	to	derive	the	respective	dispersion	
relation from	 experimental	 observations	 and	 measurements.	 This	 dispersion	 relation
determines	 the	structure	of	 the	wave	equation and	 thus characterizes	 the	dynamics	of	
the	respective	wave.	Different	wave phenomena are	thus	governed	by	different	differential
equations.		Here	we	want	to	emphasize	the	experimental	approach	to	matter	waves,	
but	before	doing	so	we	will	discuss	and	test	the	procedure	 for	other	types	of	waves,	in	
particular	water waves.

\section{Preludium}

There are a number of ways how one can motivate or even derive the wave equation
for matter waves from theoretical principles. In two recent papers Schleich, Greenberger, Kobe and Scully revisit the Schr\"odinger equation and point out the importance of the
phase dynamics and of the strong coupling between phase and amplitude \cite{Schleich2013a,Schleich2015}. In this
lecture we want to show to what extent one can obtain the Schr\"odinger equation starting
from specific experimental observations. we will not simply “postulate the classical-to-quantum rules” \cite{Schleich2013a}, but show in a consistent way how one can construct a wave equation
from experimental observations of a wave phenomenon. In principle this approach works
for any wave phenomenon and some examples are given.

Any	linear	partial	differential	equation of	spatio-temporal	variables is	solved	by	a	plane	wave ansatz 

\begin{equation}
    f(t,\vec{x}) = e^{-i\omega t + i\sum_{l=1}^{D} k_l x_l}.
\label{planewave}
\end{equation}
with $\omega$ denoting the	 temporal	 frequency	and 
$\vec{k} = \sum_{l=1}^{D} k_l \, \vec{x}_l^{(0)}$  denoting the wave vector or spatial	frequencies.	The	dimension	D can	be	one,	two	or	three,	and	the	 $\vec{x_l}^{(0)}$ represent	the	 D	 unit	 vectors	 in	 position	 space.	 Inserting	 Eq.\eqref{planewave} in	 any	 linear	 partial	 differential	
equation	yields	a	relation	between	$\omega$ and	$k_l$, l =1	to	D. In	the	standard	cases	this	results	
in	a	relation	between	$\omega$ and	 
$|\vec{k}| = k$ called	 the	dispersion	relation	 for	 this	 type	of	wave	phenomenon.	In	 this	 process	 each	 temporal	 derivative	$\frac{\partial}{\partial t}$ yields	$-i\omega$	 	 and	 each	 spatial	derivative	 	$\frac{\partial}{\partial x_l}$ yields	$ik_l.$

Conversely,	 if	 you	 have	 the	 characteristic	 dispersion	 relation	 for	 a	 particular	 type	 of		
wave	phenomenon $ g(\omega,k_1,k_2,k_3) = 0$,	 
{then you can obtain the corresponding wave equation by making the following substitutions in 
\(g(\omega, k_1, k_2, k_3)\omega(\vec{x},t) \):
\begin{itemize}
    \item[] - substitute \( i\,\frac{\partial}{\partial t} \) for \( \omega \),
    \item[] - substitute \( -i\,\frac{\partial}{\partial x_l} \) for each \( k_l \). 
\end{itemize}
}
The function $\psi(\vec{x},t)$ describes	the	wave	motion	provided	it	is	a	solution	to	the	
equation:
\begin{equation}
    g\!\left(
        i \frac{\partial}{\partial t},
        -i \frac{\partial}{\partial x_1},
        -i \frac{\partial}{\partial x_2},
        -i \frac{\partial}{\partial x_3}
      \right)\psi(\vec{x},t) = 0.
\label{gen-diff}
\end{equation}

Note,	 that	 while	 this	 procedure	 tells	 us	 the	 mathematical	 structure	 of	 the	 differential	
equation	determining	 the	wave	dynamics,	called	 the	wave	equation,	it	 tells	us	nothing	
about	 the	 physical	 nature	 of	 "the wave".	 To	 illustrate	 this	 point,	 let	 us	 take	 light	 in	
vacuum	and	apply	 the	procedure.	First	we	need	experimental	evidence	 for	 the	dispersion
relation.	This	we	get	from	astronomical	observations:	objects	on	the	sky,	changing	
with	time,	do	not	change	their	colour,	such	as	a	moon	reappearing	from	behind	Jupiter, the	varying	star	Algol	in	Perseus or	a	supernova	like	SN	1987A. This argument is attributed to Dominique François Jean Arago. see p. 225 in \citep{Lommel1875}.	This	means	that	the	
speed	of	light	is	the	same	for	all	frequencies or	$\omega = c|\vec{k}|$. Strictly	speaking,	it	is	the	group	
velocity,	which	is	constant.	This	would	imply	that	the	frequency	is	linearly	dependent	on	
the	modulus	of	$\vec{k}$ including	a	constant	offset.	Here	we	assume	that	this	offset	is	zero.	For	
the	experimental	evidence	one	has	to	measure	both	the	frequency	and	the	wavelength.	
The	former	is	facilitated	by	the	frequency	comb	technique	developed	by	Hänsch \citep{Hansch2005}	and	
Hall \citep{Hall2005}	 and	 the	 latter	 is	 routinely	 done	 with	 a	 spectrometer.	 Without	 coherent	 light	
waves	we	would	not	have	this	information.
Before	we	can	apply	 the	 transformation	used	 to	obtain	Eq.\eqref{gen-diff}	we	have	 to	get	rid	of	 the	
modulus	of $\vec{k}$.	
\\

A straight forward way to eliminate the modulus is to  square	$\omega = c|\vec{k}|$ yielding $\omega^2=c^2 (k_x^2+k_y^2+k_z^2).$ Applying	 the	 above	 transformation
we	obtain	the	familiar	wave	equation	for	light:

\begin{equation}
\bigg[ \frac{\partial^2}{\partial t^2}
      - c^{\,2}\!\left(
          \frac{\partial^{2}}{\partial x^{2}}
        + \frac{\partial^{2}}{\partial y^{2}}
        + \frac{\partial^{2}}{\partial z^{2}}
      \right)
\bigg] \Psi(t,x,y,z) = 0.
\end{equation}

In	 the	 case	 of	 light	 we	 cannot	 easily	 'see'	the	 physical	 quantity, which	 oscillates.	 As	 a	
result,	in	the	history	of	optics	one	had	to	make	assumptions	about	some	fluidum	which	
was	called	ether	which	is	the	substance	that	oscillates	and	carries	the	energy.	Needless	
to	emphasise	that	this	ether	had	to	uniformly	fill	the	whole	universe.	After	James	Clark	
Maxwell's	stroke	of	genius,	making	the	connection	between	light	and	electromagnetism,		
the	 electric	 field	 and	 the	 magnetic	 field	 were	 associated	 with	 this	 ether.	 At	 the	 time	
scientists	were	 thinking	in	 terms	of	 the	Galileo	 transformation	when	relating	mutually	
moving	coordinate	systems	and	expected	 to	be	able	 to	see	an	effect	of	 the	lab	and	 the	
Earth	 moving	 through	 this	 ether.	 Using	 a	 specially	 developed	 interferometer	 Albert	
Abraham	Michelson	showed	 first	alone	and	later	with	Edward	Williams	Morley \citep{Michel1881} and	
with	 higher	 precision	 that	 there	 was	 no	 evidence	 for	 this	 relative	motion.	 In	 order	 to	adjust	the	theory	to	cope	with	this	null	result	George	Francis	FitzGerald \citep{Fitz1889}	proposed	to	
replace	the	Galileo	transformation	by	a	new	transformation	which	a	few	years	later	was	
also	 derived	 by	Hendrik	Antoon	 Lorentz \citep{Lorentz1892}.	The	 Lorentz-transformation,	as	it	is	 often	
called,	 introduced	 a	 Lorentz-invariant	 ether.	 Maxwell's	 equations	 for	 empty	 space	
survived	 because	 they	 were	 curiously	 enough	 invariant	 under	 the	 Lorentz	
transformation.	It	was	when	Einstein	showed	that	the	Lorentz	transformation	is	a	direct	
consequence	of	assuming	that	the	speed	of	light	in	empty	space	is	the	same	in	all	inertial	
co-ordinate	 systems	 and	 independent	 of	 their	 relative	 velocity,	 that	 a	 paradigm	 shift	
happened	and	 the	concept	of	 the	ether	was	abandoned.	Since	 then	 the	name	ether	has	
this	old-fashioned	and	backward	connotation.	However, our	modern	concept	of	the	vacuum	is	
not	 that	 of	 empty	 space,	 but	 rather	 of	 a	 space filled	 with	 vacuum	 fluctuations.	 The	
vacuum	– acting	much	like	a	dielectric	would	do	– screens	the	bare	point	charge	of	e.g.	
the	 electron \citep{Gott1986},\citep{Hawton2016a},	 an	 effect	 called	 vacuum	 polarization, to	 give	 rise	 to	 the	 elementary	
charge	observed	in	low	energy	experiments. At	higher	energy	collisions	between	point	
charge	particles	the	screening	reduces \citep{Hogan2000},\citep{Hawton2016b} giving	rise	to	a	modification	of	the	effective	
charge	and	equivalently	 to	 the	 so-called	 running	 of	 the	 fine-structure	 constant	 (for	all	
this,	see	also	the	"Further readings" section, second item).	As	a	result one	might	today	think	again	of	
ether,	but	now	it	is	a	modern	Lorentz-invariant	quantum	ether	the	properties	of	which	
are	determined	 by	 the	 different	 types	 of	 charged	 elementary	 particles	 responsible	 for	
the	 vacuum	 fluctuations.	And	 these	properties	determine	 the	parameters	appearing	in	
Maxwell's	 equations	 giving	 rise	 to	 an	 interesting	 connection	 between	 high-energy	
physics	 and low-energy	 optics	 (see	 "Further readings" section).	 Anyway,	 the	 purpose	 of	 the present section is	 to	 underline	 and	 emphasize	 that dispersion	 relations	 tell us	 little	 if	
anything	 about	 the	 physical	 nature	 of	 a	 wave,	 it	 tells	 us	 only	 about	 the	 mathematical	
structure	of	the	wave	equation.

\section{Water	waves}
Water	waves	are	an	interesting	example,	as	we	will	see.	The	 type	we	will	discuss	here	
are	the	\emph{deep	water	waves},	not	affected	by	a	finite	water	depth,	and	waves	of	wavelength	
longer	than	a	few	centimetres,	for	which	surface	tension	can	be	neglected.	The	properties
of	this	special	type	of	water	wave	is	governed	by	gravitation.

{ Deep-water waves are \emph{dispersive}, meaning that different wavelengths travel with different phase velocities. 
This property will be reflected in the dispersion relation determined below.}

We	 begin	 by	 experimentally	 determining the	 dispersion	 relation.	 A	 straightforward	
experiment	 uses	 a	 wave	 machine.	 A	 movable	 wall	 in	 a	 large	 enough	 basin	 moved	
forward	and	backward	with	a	 time	period	of	T.	The	medium	responds	by	developing	a	
wave	with	a	particular	wavelength	$\lambda$.	The	experimental	finding	is	that~\citep{Lamb1932}

\begin{equation}
    \lambda \propto T^2.
\label{lambda-T-water}
\end{equation}

\begin{figure}[h!]
\centering
\includegraphics[scale=1.7]{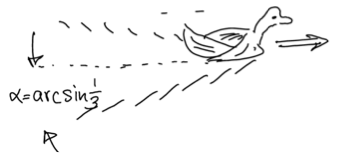}
\caption{Wave	 pattern	 behind	 a	
duck	travelling	on water}
\label{duck}
\end{figure}

This result	 is	 independent of	 amplitude because	 both the inertial property  and  the gravitational	restoring	force	are	proportional to	the	same	mass	in	a	single	hump	of	the	wave.

{There are other ways to obtain the same relation. One may simply stand at a pond and throw stones into the water while observing the developing pattern~\citep{Leuchs1991}, possibly recording it on video.}
For	a	particular	size	and	weight	of	the	stone	you	will	get	a specific	pair	of	values	for	the	temporal and	the	spatial	period.	Using	stones	of	different	
size	and	weight	you	can	get	different	pairs	of	values.	Looking	closely	one	sees	 that	 the	
circular	wave	 pattern	 created	after	a	 single	 throw	 of	a	 stone	 develops	a	 chirp,	 further	
out	 the	 periods	 become	 longer.	 Thus	 the	 video	 taken	 after	 a	 single	 throw	 with	 one	
particular	stone	will	give	you	different	pairs	of	values.	All	combined	one	again	retrieves $\lambda \propto T^2.$

The	 last	 way	 I	 want	 to	 mention	 briefly involves	
analysing	 the	wave	pattern	behind	any	 object	moving	
fast	 enough	 through	 water.	 A	 duck	 at	 typical {swimming} speed is {sufficiently}	 fast.	Behind	 these	 objects	 develops	a
conical	 wave	 pattern	 the	 opening	 angle	 of	 which	 is	
surprisingly	independent	of	the	speed (Fig.	\ref{duck}).	Already	
William	 Thomson,	 better	 known	 as	 Lord	 Kelvin,	
determined	this	opening	angle	to	be	$\theta = 2\arcsin(1/3) \approx 38.94^\circ$~\citep{Kelvin1887}.	This result is obtained when assuming the	group	velocity	of waves in deep water is	half as	large	as	 the	phase velocity:

\begin{equation}
     \frac{\partial\omega}{\partial k}=\omega/2k.
\end{equation}

Integrating	 this	equation	 yields	again	 the	relation	shown	in	Eq.\eqref{lambda-T-water}.This	corresponds	 to	
the	 dispersion	 relationship $\omega ^2 \propto  k.$	 The	 proportionality	 constant	 must	 have	 the	
dimension	 of	acceleration.	Considering	 that	 these	waves	are	driven	 by	gravitation	 one	
might	guess	that	it	should	be	the	gravitational	acceleration	on	Earth,	which	is	$g$.	The	full	
dispersion	relation {for deep-water gravity waves is therefore} (see p. 354 in \cite{crawford1965}):
 \begin{equation}
      \omega ^2 = gk.
 \end{equation}
{Using the relations $\omega = 2\pi/T$ and $k = 2\pi/\lambda$, one directly verifies that 
$\omega^2 = gk$ implies $\lambda \propto T^2$.}

 \subsection{Wave	equation	for	water	waves}
 {In this case we again encounter }
the	 complication	 that	 the	modulus	 of	 the	 wave	 vector $\vec{k}$ appears.	
Unlike	 in	 the	 case	 of	 light	 discussed	 in	 the preludium	 it	 makes	 sense	 to	 restrict	 the	
discussion	 to	 one	 dimension	 of	$\vec{k}$ .	 This	 already	 reveals	 some	 non-trivial	 properties	 of	
these	special	types	of	waves.	The	corresponding	wave	equation	is (see also \cite{Weisman2017a}):
\begin{equation}
    \frac{\partial^2}{\partial t^2} \Psi (t,x) = gi \frac{\partial}{\partial x} \Psi (t,x).
\label{compl-diff-water}
\end{equation}
 We	 note	 two {important }points:	
 
{First, in this particular case we know from the observation the physical nature of the wave that it is the vertical displacement (height) of the water surface that oscillates.}

Secondly, the	 differential	 equation	 contains	 the	 imaginary	 unit {$i$}.	  
{This seems to imply that the solution must in general be complex-valued.}
But	how	can	{a complex solution describe the real, measurable} height	of	the	water	surface?	

Let	us	first	address	this latter	point.	We	can	eliminate	
the	{explicit} imaginary	unit	by	writing	 the	wave	 function	as	 the	sum	of	a	real and	an	imaginary part $\Psi = \Psi_r + i\Psi_i.$	Inserting	this	into	Eq.\eqref{compl-diff-water}	the	real	terms	on	the	left	side	have	to	match	
the	real	 term	on	 the	right	and	 the	same	holds	 for	 the	imaginary	 terms.	This	yields	 two	
real	valued	coupled	differential	equations:

\begin{equation}
    \begin{array}{llll}
        \frac{\partial^2}{\partial t^2} \Psi_r(t,x) = -g \frac{\partial}{\partial x}\Psi_i(t,x), \\[6pt]
        \frac{\partial^2}{\partial t^2} \Psi_i(t,x) = \;\;\, g\frac{\partial}{\partial x}\Psi_r(t,x).
    \end{array}
\label{real-diff-water}
\end{equation}

We see that the imaginary unit in Eq.\eqref{compl-diff-water} simply means that the differential equation describes the combined dynamics of two real functions. In the two equations in \eqref{real-diff-water} the roles of   $\Psi_r$ and  $\Psi_i$  are interchanged and there is a difference in a sign {change}.  The set of Eqs.\eqref{real-diff-water} is solved by

\begin{equation}
    \begin{array}{ll}
        \Psi_r (t,x) = a  \sin(\omega t-kx) \\\Psi_i (t,x) = a  \cos(\omega t-kx).
    \end{array}             
\label{sol-diff-water}
\end{equation}

\begin{figure}
\makebox[\linewidth]{
\includegraphics[width=\linewidth]{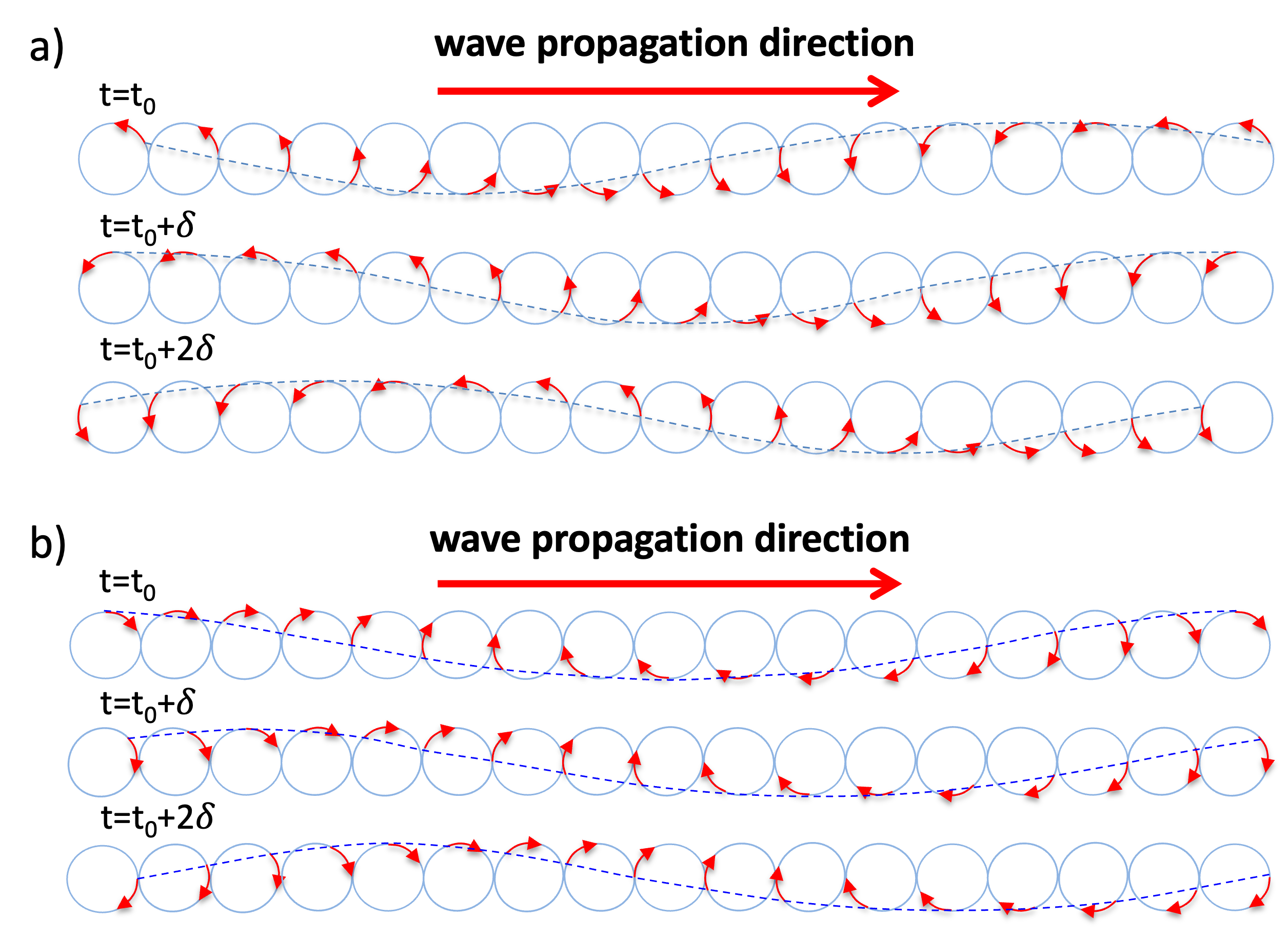}}
\caption{Propagation of deep-water surface waves is the result of a circular motion of the surface molecules. The dotted lines represent the surface. Formally, two different solutions are possible. (a) Variant 1: the individual molecules in the crest are moving in the opposite direction to the crest as indicated by the red arrows. The same is true - to a lesser extent - for the molecules close to the surface (details of the collective motion not shown). (b) Variant 2: the individual molecules in the crest are moving along with the crest.  Only (b) is physically relevant. See text for details.}
\label{crest forward}
\end{figure}

For positive $k$, Eq.\eqref{sol-diff-water} describes a wave propagating in the positive $x$-direction. We already mentioned that observation reveals that the height of the water surface is oscillating. 

But the height is only one real variable. What is then the second real variable, i. e. the other one of the two functions? Taking	a	
closer	 look	 at	 water	waves	 we	 notice	 that	 the	surface	 is	 not	 only	 going	up	 and	 down	 but	 also	 to	
and	 fro.	 
{We observe this whenever an object is 
floating in the water and follows an up–down and to–and–fro motion.
}

A	particular	point	on	the	 surface	is	 obviously	making	a	circular	motion	properly	described	by	Eqs.\eqref{sol-diff-water}.	So	we	
found	 that	 the	 two	 functions $\Psi_r$  and $\Psi_i$ describe	 the	vertical	and	longitudinal	periodic motions	of	the	positions	of	surface	molecules.	

However, our approach does not tell us whether the height of the surface is described by the real or the imaginary part of the wave function. So we have two choices resulting in two different wave patterns. 

In Fig.~\ref{crest forward}(a) the water molecules at the crest move against the propagation direction of the crest and the crest is wider than the trough. This solution is, however, not compatible with the observation in nature\footnote{This is a valid mathematical model compatible with Eq.\eqref{sol-diff-water}, http://demonstrations.
wolfram.com/MotionOfParticlesInOceanWaves/. However, Yuliya Troitskaya of IAP-RAS
pointed out to me that for the solution shown in Fig. \ref{crest forward} a the amplitude of the circular motion grows
exponentially with distance from the surface, which is unphysical.}

Fig.~\ref{crest forward}(b) shows the other possibility compatible with Eq.\eqref{sol-diff-water}, the water molecules at the crest move along with the crest and the crest is more peaked. For a video of this motion see the footnote\footnote{https://www.youtube.com/watch?v=NShUBfJQEHk . The video shows the physically viable solution. And note, that this circular motion holds for all deep water waves, not only wind generated ones.}. Inspection of water waves in nature shows indeed that Fig. \ref{crest forward}(b) shows the proper motion of water waves. For	a	full	
mathematical	treatment	see	e.g.	Horace	Lamb \cite{Lamb1932}.

{Everybody enjoying the waves near a beach has experienced the power of the wave crest when being pushed toward the shore.}
This is	 because one floats	 on	 a	tilted	
surface. The	 trick	 is	 to	 dive	 under	 the	 crest – because	 of	 the	 motion	 of	 the	 water	molecules	 on	 and	 underneath	 the	 crest	 is	 opposite	 to	 the	 wave	 propagation	 and	 one swiftly	 reappears	in	 the	 next	 valley.	One	 can	 thus	 exploit	 this	 special	 wave	 pattern	in	both	directions:	diving	under	the	crest	when	going	into	the	sea	and	body	surfing	on	the	slope	preceding	the	crest	when	coming	back.	

It	is	quite	remarkable	that	all	this	is	made possible by	 the	 occurrence	 of	 the	 imaginary	 unit	 in	 Eq.\eqref{compl-diff-water},  giving	 rise	 to	 the	 circular	motion. 

Note	 that	 recently,	when	analysing	 the	wake	 field	of	 fast	boats, Rabaud	and	Moisy	\citep{Rabaud2013}	
reported	on	some	discrepancies with	Lord	Kelvin's	postulate that	 the	wake field	angle	is	
constant. The observed reduction of the wake field angle is attributed to {the fact that sufficiently} fast,
small boats do not necessarily excite any water waves with a phase velocity at or near to
the boat velocity.

\section{Matter waves}
With	 this	 digression	 to	 water	 waves	 we	 are {now} well	 prepared	 to	 tackle	 the	 problem	 of	matter	waves.	Again	we	start	by	looking	for	some	experimental	evidence	for	the	dispersion relation,	i.e.\ for	a	wave	phenomenon	in	nature	associated	with	matter	waves	hoping	to	 extract	 the	 dispersion	 relation	 for	matter	 waves.	 The	 obvious	 showcases	 are	interference patterns,	which	 we	 recognize	as	 such,	 when	 a	
matter	 wave	 is	 diffracted off	 a	 small	structure.	 It	 goes	without	 saying	 that	this	 structure	 must	have	 a	 dimension	somewhat	 comparably to	 the	 wavelength of	 the	 matter	waves	 in	 order	 to	see	 a	 pronounced	effect.	
{This requirement is directly analogous to optical diffraction, where slit or lattice 
spacings must be comparable to the wavelength of light.}

\begin{figure}[h!]
\centering
\includegraphics[scale=1]{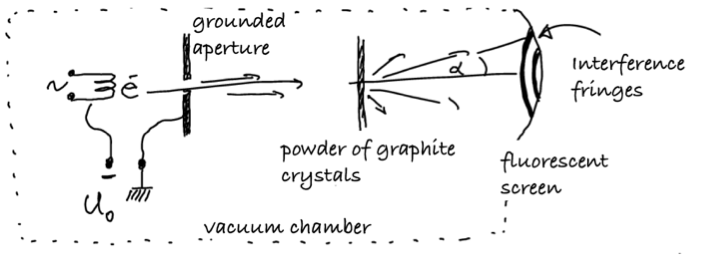}
\caption{Sketch	of	an	apparatus	for	studying	the	diffraction	of	electrons	
after	impinging	on	 graphite.	The	 angle	$\alpha$ is	 proportional	 to	 the	wavelength,
 as	 we	 know	 from	 analogous	 scenarios	 in	 optics.	 This	 angle	
changes	when	the	electron	acceleration	voltage	$U_0$ is	varied}
\label{exp-e-diffraction}
\end{figure}

We choose diffraction of electrons by randomly oriented graphite crystals 
(Fig.\ref{exp-e-diffraction}). For first diffraction from single crystal see~\cite{Davisson1927} and for early work on diffraction from polycrystalline substance see~\cite{Thomson1928}. In a vacuum chamber, electrons emitted by a hot wire at 
high negative potential $-U_0$ accelerate toward a grounded metal aperture. 
Electrons passing through the central aperture hole emerge as an expanding 
electron beam with an average velocity given by
\begin{equation}
    u^{2} = \frac{2eU_0}{m}.
\end{equation}
The mass $m$ and charge $-e$ of the electron are fixed quantities. The electron 
beam hits a thin target made of graphite powder, scattering the electrons toward 
a fluorescent screen. The	fluorescence	intensity	at	a	particular	point	on	the	screen	indicates	the	flux	of	
electron impinging	 at	 this	 point.	 What	 one	 observes	 is	 a	 pattern	 of	 concentric	 rings	reminiscent	 of	 the	 rings	 around	 a	 bright	 yellow	 streetlight when	 looking	 through	the	fogged	window	of	a	bus	at	wintertime.	

{The halo occasionally seen around the Moon is the same optical phenomenon, except that a monodisperse layer of ice crystals in the upper atmosphere takes over the role of the fogged glass.
Likewise, the effect observed with fogged glass becomes more pronounced and more colorful the more monodisperse the size distribution of the fine droplets on the glass is.
The angle under which a ring is observed is proportional to the wavelength — blue light appears at a larger angle than red light.}

We	do	 not	 see	 the	matter	 wave	 directly,	 but	 we	 see	 something,	 which	is	 very	 close	 to	 an	interference	 pattern	 we	 know.	 Hence	 we	 proceed	 with	 the	 assumption	 that	 electrons	behave	 like	 waves	 under	 this	 condition.	 Based	 on	 this	 analogy	 we	 conclude	 that	 also	here	the	angle $\alpha$ is	proportional	to	the	wavelength	of	the	electrons for	small	angles	 for which	the	$\sin$	function	can	be	replaced	by	its	argument.	

Next	we	vary	the	acceleration	voltage	$U_0$  and	find	that	this	changes	also	the	angle	under	which	the	rings	appear	on	the	screen.	A more quantitative measurement	yields:	
\begin{equation}
    \alpha \propto U_0^{-1/2} = \frac{\sqrt {2e/m}}{u}. 
\label{e-rings}
\end{equation}

{When thinking of electrons as waves we naturally think of more or less localized wave packets. }
In	wave	language	the	group	velocity	$ u =  \frac{\partial\omega}{\partial k}$  gives	the	speed	
of	the	wave	packet.	Using	in	addition	$\alpha \propto \lambda$,	we	can	transform	Eq.\eqref{e-rings}	into:

\begin{equation}
\begin{array}{ccc}
{\lambda \propto (\frac{\partial\omega}{\partial k})^{-1}}&{or}&{\frac{\partial\omega}{\partial k} = A k}.      
\end{array}
\label{disp-matter}
\end{equation}
where	the	constant	$A$	turns	out	to	be	 $\hbar / m$, see footnote\footnote{When	repeating	the	experiment	with	particles	of	other	mass,	one	notices	that	the	wavelength	depends	not	only	on	the	velocity	but	also	on	the	mass,	such	that	the	linear	momentum	of	the	particle is	inversely	
proportional	to	the	wavelength.	This	proportionality	constant	is	a	fundamental	constant:	Planck's	
constant  $\hbar$. It is in	principle	possible to	determine  $\hbar$ from	diffraction	experiments	of	this	kind.}. This translates	to	the	integral	
form
\begin{equation}
\begin{array}{cccc}
{\int d\omega = \frac{\hbar}{m}\int k dk}&{\mathrm{and}}&{\mathrm{to}}&{\omega =\frac{\hbar}{2m}k^2 + C}.
\end{array}
\end{equation}

This	is	the	dispersion	relation	for	matter	waves	we	were looking	for. {The constant $C$ is	an integration	
constant}.	It	is	not	a	function	of	$k$ or $\omega$, but may well be a function of other parameters such as position and time.
\subsection{Wave	equation	for	matter	waves}
In	 Eq.\eqref{disp-matter},	it	is	 the	 square	 of	 the	wave	 vector,	which	enters.	Therefore	we	 can	 directly	
proceed	with transforming	$k$ and $\omega$ into	 the	corresponding	differential	operators,	 yielding:
\begin{equation}
 i\frac{\partial}{\partial t} \Psi (t,\vec{x}) = -\frac{\hbar}{2m}\Delta  \Psi (t,\vec{x}) +C(t,\vec{x})\Psi (t,\vec{x}).
\label{diff-matter}
\end{equation}
Multiplying	 Eq.\eqref{diff-matter} with $\hbar$ yields	 the	 Schr\"odinger	 equation (see	 p.	 112	 in	 \cite{Schröd1926}) and	 $C$	
allows	 for	accommodating	the	 potential $V$,	which	may	 depend	 on	 the	 spatial	 but	also on
the	temporal	coordinates. 
\begin{equation}
  i\hbar\frac{\partial}{\partial t} \Psi (t,\vec{x}) =  -\frac{\hbar ^2}{2m}\Delta  \Psi (t,\vec{x}) + V(t,\vec{x}) \Psi (t,\vec{x}).
\label{SchröGl}
\end{equation}

 \section{Final	remark}
 We	thus	achieved	our	goal of deriving	the	wave	equation	for	matter	waves	and	could {in principle} stop here.	But	the	journey	through	the	empirical	derivation	of	the	various	wave	equations	for	
the	different	types	of	wave	phenomena	in	physics	gives	also	room	for	speculation.	

In	the	case	 of	 the	water	waves, the	 asymmetry	 between	 the	 temporal	 and	 spatial	 differential	operators	gave	rise	to	the {appearance of the} imaginary	unit	in	the	wave	equation,	which	in	turn	meant	that	
the	wave	equation	can	be	written	as	two	coupled	real	differential	equations. Indeed, we	 found	 that	 the	 water	 waves	 are	 associated	 with	  a two-dimensional motion.

Now,	 in	 the	 case	 of	
matter	waves,	we	have a similar situation	and	could	write Eq.\eqref{SchröGl}	as:

\begin{equation}
\left\{
\begin{array}{ll}
\hbar\,\frac{\partial}{\partial t} \Psi_i(t,\vec{x})
   = \;\;\frac{\hbar^2}{2m}\,\Delta \Psi_i(t,\vec{x})
     - V(t,\vec{x})\,\Psi_i(t,\vec{x}), \\[10pt]

\hbar\,\frac{\partial}{\partial t} \Psi_r(t,\vec{x})
   = -\frac{\hbar^2}{2m}\,\Delta \Psi_r(t,\vec{x})
     + V(t,\vec{x})\,\Psi_r(t,\vec{x}).
\end{array}
\right.
\end{equation}
with	 real	 functions	$\Psi_r$ and	$\Psi_i$ .	 
{So	 far	 we	 have	 a	 physical	 interpretation	 only	 for	 the	
squared	 modulus	 of	 the	 full	 wave	 function	$\Psi ^2 = \Psi_r ^2 + \Psi_i ^2$ and	 one	 might	 speculate	
whether	novel	experiments	might	bring	to	light	more	physical	insight.}
{Recently,
in a most remarkable paper, Schleich, Greenberger, Kobe, and Scully \cite{Schleich2015} also address this
very question. But they followed a different approach. Instead of writing $\Psi (\vec r, t) = \Psi_r (\vec r, t)+ i\Psi_i (\vec r, t)$
they used an amplitude and a phase function, $\Psi(\vec r, t) = A(\vec r, t)e^{i\theta (\vec r, t)}$, with the following interpretation: The square of the amplitude represents the
spatial density ( or rather the position ) and the gradient of the phase represents the
momentum. This provides insight into the interpretation of the wave function, although information about amplitude and phase is also available from wave functions solving a single real valued differential equation. The physical significance of the two dimensional aspect of the wave equation for matter waves is still an open question.}

Here	we	can	only conclude	by	emphasizing	 that	 the	empirical	route	 to	 the	mathematical	structure	of	 the wave	 equation	 describing	 a	 particular	 wave	 phenomenon	 gives	 by	 itself	 no	 clue whatsoever	 about	 the	 physical	 nature	 of	 the	 wave	 function	 which	 solves	 the	 wave	
equation. Experiments	 with	 coherent	 matter	 waves	 are	 now	 becoming	 routine	 and might	 give some	 new	 insight	 at	 some	 point,	 much	 like	 the	 availability	 of	 coherent electromagnetic	waves (such as radio waves) has	helped	our	understanding	of	this	phenomenon.
 
 
\section{Further reading}

Regarding the two other lectures given by Gerd Leuchs in 2016, we would like to draw attention to 
manuscripts available on \texttt{arXiv.org}:

\begin{enumerate}
  \item \emph{Time reversal symmetry -- a powerful tool in quantum optics},\\
        arXiv:\,1205.1374 and arXiv:\,1309.6167.
  \item \emph{Evidence at low energy for high energy particles},\\
        arXiv:\,1612.03394 and arXiv:\,1612.06263.\\
        (The latest manuscript on this topic: arXiv:\,2507.19569, added 22Nov2025)
\end{enumerate}

\section{Acknowledgement}

 thank Shadi Sorayya and Ankan Bag for helping preparing the manuscript.


\bibliographystyle{plain}
\bibliography{ref}

@book{Lommel1875,
  author    = {Eugen Lommel},
  title     = {The Nature of Light: With a General Account of Physical Optics},
  year      = {1875},
  publisher = {Henry S.~King \& Co.},
  address   = {London},
  note      = {This argument is attributed to Dominique Fran{\c{c}}ois Jean Arago; see p.~225.},
}

@manual{Hansch2005,
  author = {H{\"a}nsch, T.~W.},
  title  = {Nobel Lecture},
  year   = {2005},
  note   = {Available at: \url{https://www.nobelprize.org/prizes/physics/2005/hansch/lecture/}}
}

@manual{Hall2005,
  author = {Hall, J.~L.},
  title  = {Nobel Lecture},
  year   = {2005},
  note   = {Available at: \url{https://www.nobelprize.org/prizes/physics/2005/hall/lecture/}}
}

@article{Michel1881,
  author  = {Michelson, A.~A. and Morley, E.~W.},
  title   = {On the Relative Motion of the Earth and the Luminiferous Ether},
  journal = {American Journal of Science},
  volume  = {34},
  pages   = {333--345},
  year    = {1881}
}

@article{Fitz1889,
  author  = {FitzGerald, G.~F.},
  title   = {The Ether and the Earth's Atmosphere},
  journal = {Science},
  volume  = {13},
  pages   = {390},
  year    = {1889}
}

@article{Lorentz1892,
  author  = {Lorentz, H.~A.},
  title   = {Zittingsverlag Akad. v. Wet.},
  journal = {Verslagen van de Afdeeling Natuurkunde der Koninklijke Akademie van Wetenschappen},
  volume  = {1},
  pages   = {74--79},
  year    = {1892}
}

@book{Gott1986,
  author    = {Gottfried, K. and Weisskopf, V.~F.},
  title     = {Concepts of Particle Physics},
  volume    = {II},
  publisher = {Oxford University Press},
  address   = {New York},
  year      = {1986}
}

@misc{Hawton2016a,
  author = {Hawton, M. and S{\'a}nchez-Soto, L.~L. and Leuchs, G.},
  title  = {The Linear Optical Response of the Quantum Vacuum},
  year   = {2016},
  eprint = {1612.06263},
  archivePrefix = {arXiv}
}

@article{Hogan2000,
  author  = {Hogan, C.~J.},
  title   = {Why the Universe Is Just So},
  journal = {Reviews of Modern Physics},
  volume  = {72},
  pages   = {1149--1161},
  year    = {2000}
}

@misc{Hawton2016b,
  author = {Hawton, M. and S{\'a}nchez-Soto, L.~L. and Leuchs, G.},
  title  = {Quantum Field Theory and Classical Optics: Determining the Fine Structure Constant},
  year   = {2016},
  eprint = {1612.03394},
  archivePrefix = {arXiv}
}

@book{Lamb1932,
  author    = {Lamb, Horace},
  title     = {Hydrodynamics},
  publisher = {Dover Publications},
  address   = {New York},
  year      = {1945},
  note      = {Originally published by Cambridge University Press, 6th ed.\ (1932)}
}

@article{Leuchs1991,
  author  = {Leuchs, G. and Deissinger, E.},
  title   = {Article in P.M. Magazin},
  journal = {P.M. Magazin},
  pages   = {70},
  year    = {1991},
  note    = {Issue 9, see p.~74}
}

@article{Kelvin1887,
  author  = {Thomson, W.},
  title   = {On Ship Waves},
  journal = {Proceedings of the Institution of Mechanical Engineers},
  volume  = {38},
  pages   = {409--434},
  year    = {1887}
}

@article{Rabaud2013,
  author  = {Rabaud, M. and Moisy, F.},
  title   = {Ship Wakes: Kelvin or Mach Angle?},
  journal = {Physical Review Letters},
  volume  = {110},
  pages   = {214503},
  year    = {2013}
}

@article{Thomson1928,
  author  = {Thomson, G.~P.},
  title   = {Diffraction of Cathode Rays by Thin Films},
  journal = {Proceedings of the Royal Society of London A},
  volume  = {117},
  pages   = {600--609},
  year    = {1928}
}

@article{Davisson1927,
  author  = {Davisson, C. and Germer, L.~H.},
  title   = {Reflection of Electrons by a Crystal of Nickel},
  journal = {Physical Review},
  volume  = {30},
  pages   = {705--740},
  year    = {1927}
}

@article{Schröd1926,
    author  = {Schr{\"o}dinger, Erwin},
    title   = {Über die Eigenwerttheorie},
    journal = {Annalen der Physik},
    volume  = {81},
    pages   = {109--139},
    year    = {1926},
}

@article{Schleich2013a,
	author = {Wolfgang P. Schleich and Daniel M. Greenberger and Donald H. Kobe and Marlan O. Scully},
	doi = {10.1073/pnas.1302475110},
	eprint = {https://www.pnas.org/doi/pdf/10.1073/pnas.1302475110},
	journal = {Proceedings of the National Academy of Sciences},
	number = {14},
	pages = {5374-5379},
	title = {Schr{\"o}dinger equation revisited},
	url = {https://www.pnas.org/doi/abs/10.1073/pnas.1302475110},
	volume = {110},
	year = {2013}}

@article{Weisman2017a,
  title = {Diffractive Focusing of Waves in Time and in Space},
  author = {Weisman, Dror and Fu, Shenhe and Gon\ifmmode \mbox{\c{c}}\else \c{c}\fi{}alves, Manuel and Shemer, Lev and Zhou, Jianying and Schleich, Wolfgang P. and Arie, Ady},
  journal = {Phys. Rev. Lett.},
  volume = {118},
  issue = {15},
  pages = {154301},
  numpages = {6},
  year = {2017},
  month = {Apr},
  publisher = {American Physical Society},
  doi = {10.1103/PhysRevLett.118.154301},
  url = {https://link.aps.org/doi/10.1103/PhysRevLett.118.154301}
}

@article{Schleich2015,
	author = {Schleich, W P and Greenberger, D M and Kobe, D H and Scully, M O},
	doi = {10.1088/0031-8949/90/10/108009},
	journal = {Physica Scripta},
	month = {sep},
	number = {10},
	pages = {108009},
	publisher = {IOP Publishing},
	title = {A wave equation interpolating between classical and quantum mechanics*},
	url = {https://doi.org/10.1088/0031-8949/90/10/108009},
	volume = {90},
	year = {2015}}

@book{crawford1965,
  author    = {Crawford, Frank S. Jr.},
  title     = {Waves - Berkeley Physics Course - volume 3},
  volume    = {3},
  publisher = {McGraw-Hill},
  address   = {New York},
  year      = {1965}
}
\end{document}